\documentclass[%
 aip,
 amsmath,amssymb,
 reprint,%
]{revtex4-1}

\usepackage{graphicx}
\usepackage{dcolumn}
\usepackage{bm}
\pdfoutput=1
\usepackage[dvipsnames]{xcolor}
\usepackage[utf8]{inputenc}
\usepackage[T1]{fontenc}
\usepackage{mathptmx}
\DeclareMathAlphabet{\mathcal}{OMS}{cmsy}{m}{n}
\usepackage{etoolbox}
\usepackage{amsthm}
\usepackage{amsmath}
\usepackage{caption}
\usepackage{subcaption}
\usepackage{physics}
\usepackage{float}
\usepackage{hyperref}
\usepackage[normalem]{ulem}
\usepackage{bm}

\newcommand{\x}{\mathbf{x}}
\newcommand{\F}{\mathbf{F}}
\newcommand{\s}{\mathbf{s}}
\renewcommand{\H}{\mathbf{H}}
\newcommand{\deltaeta}{\bm{\delta\eta}}
\newcommand{\deltaxi}{\bm{\delta\xi}}
\newcommand{\deltax}{\bm{\delta \x}}

\usepackage{xr}

\newtheorem{proposition}{Proposition}

\newtheorem*{remark*}{Remark}
\makeatletter

\makeatletter
\def\@email#1#2{%
 \endgroup
 \patchcmd{\titleblock@produce}
  {\frontmatter@RRAPformat}
  {\frontmatter@RRAPformat{\produce@RRAP{*#1\href{mailto:#2}{#2}}}\frontmatter@RRAPformat}
  {}{}
}%
\makeatother

\begin{document}

\preprint{AIP/123-QED}

\title{Symmetry-induced activity patterns of active inactive clusters in complex networks}
\author{Anil Kumar}
\email{kumaranil@iisertvm.ac.in}
\affiliation{
School of Physics, Indian Institute of Science Education and Research, Thiruvananthapuram 695 551, Kerala, India
}%
\author{V. K. Chandrasekar}
\affiliation{
Centre for Nonlinear Science \& Engineering, School of Electrical \& Electronics Engineering, SASTRA University, Thanjavur- 613 401, India.
}%

\author{D. V. Senthilkumar}
\email{skumar@iisertvm.ac.in}
\affiliation{
School of Physics, Indian Institute of Science Education and Research, Thiruvananthapuram 695 551, Kerala, India
}%

\date{\today}

\begin{abstract}

Synchrony patterns characterize network states in which nodes organize into clusters based on their synchronized dynamics. The synchronized clusters may further exhibit either active or inactive states. The simultaneous invariance of active and inactive clusters of synchronized nodes poses a dynamical constraint because fluctuations from active clusters must cancel out for a desired cluster to be inactive. By exploiting permutation symmetries in the network structure and choosing dynamics on top such that internal dynamics and coupling functions are odd functions in the phase space, we demonstrate that this combination of structure and dynamics exhibits stable invariant patterns composed of coexisting active and inactive clusters. The symmetries in a network generate active clusters that are in antisynchrony with each other, resulting in the cancellation of fluctuations for clusters connected with these antisynchronous clusters. We use full network symmetries to obtain synchronized clusters, while quotient network symmetries are used to find coexisting active-inactive states of clusters.
We show that as the coupling between nodes changes, active clusters lose their activity at different coupling values, and the network transitions from one activity pattern to another. Numerical simulations are presented for networks of Van der Pol and Stuart-Landau oscillators. Finally, we extend the master stability framework to these patterns and provide stability conditions for their existence.  

\end{abstract}

\maketitle

\begin{quotation}

A synchrony pattern describes a network state in which nodes are partitioned into clusters whose elements evolve identically in time. While synchronized clusters are typically associated with oscillatory dynamics, they may also remain static, forming inactive states. We refer to the collective configuration of all such clusters, where some may be active and others inactive, as an activity pattern. The invariance of inactive clusters requires that fluctuations originating from neighboring clusters cancel each other exactly. Here, we show that such cancellation can arise from permutation symmetries of the network structure when the node dynamics are governed by functions that are odd in phase space. In this setting, symmetry-related nodes adopt antisynchronized configurations, whereby their opposing fluctuations cancel and maintain the invariance of inactive clusters connected to them. We generate these antisynchronized clusters from the permutation symmetries of quotient networks. Some of the invariant patterns may be transient, so a method to find their existence for all times is also presented. We show that, as the coupling between nodes varies, active clusters lose their oscillatory activity, leading the network to transition between different activity patterns. Finally, we derive a unified stability theory that determines the stability of synchronized, antisynchronized, and inactive cluster states, thereby providing a single theoretical framework for understanding the emergence and transitions of such patterns in coupled dynamical networks.
\end{quotation}

\section{Introduction} 
Investigations on synchronization in networks of coupled dynamical systems have been an evergreen topic for more than half a century~\cite{Pikovsky2001synchronization, arenas2008} due to its wide range of applications in diverse areas of science and technology~\cite{Pikovsky2001synchronization, arenas2008, TANG2014synch_review, BOCCALETTI2006complexnet}. Nevertheless, recent investigations are centered on cluster synchronization or partial synchronization  to unravel the underlying
dynamical mechanisms of intriguing macroscopic collective states.      
Cluster synchronization refers to  two or more synchronized groups, such as phase clusters~\cite{Berner2020birth}, frequency clusters~\cite{Berner2019multiclusters,Thamizharasan2022exotic}, hierarchical clusters~\cite{Berner2019hierarchical}, 
antipodal states~\cite{Berner2020birth,Berner2019multiclusters,Thamizharasan2022exotic},  bump frequency clusters~\cite{Thamizharasan2024habbian}, forced entrained clusters~\cite{Thamizharasan2024habbian}, etc., that constitute the entire network. 
Partial synchronization refers to coexisting domains of coherent and incoherent nodes of a network, such as chimera states~\cite{Abrams2004chimeras, bollt2023fractal, parastesh2021chimeras, Majhi2019chimera}, bump states~\cite{Thamizharasan2024habbian}, splay and splay cluster states~\cite{Thamizharasan2024habbian}, etc. These synchronous patterns have striking similarities with the spatiotemporal patterns
observed in neuronal networks~\cite{RATTENBORG2000behavioral, Bansal2019cognitive, bollt2023fractal}, chemical oscillator networks \cite{Totz2015phaselag}, power grid networks \cite{motter2013spontaneous}, and reveal a great deal of insight about their functional behavior~\cite{RATTENBORG2000behavioral, Bansal2019cognitive, bollt2023fractal, Totz2015phaselag, motter2013spontaneous}. 
Quite often, the underlying structure of a network has been shown to play a crucial role in the manifestation of synchrony patterns~\cite{Gardenes2007pathsto}.
In particular, the prevailing symmetry of a network has been
found to be the source of partial synchrony patterns~\cite{pecora2014cluster}. In addition, external equitable partitions \cite{schaub2016graph} or balanced node coloring \cite{Stewart2003symmetry} have also been shown to be the basis for coupling-induced synchronized clusters. 
Theoretical frameworks,
based on the tools of computational group theory to reveal the hidden symmetries of networks, have been introduced in recent years to determine the stability of cluster synchronization in complex networks with symmetries~\cite{pecora2014cluster}, in multilayer networks~\cite{della2020symmetries}, and in directed networks~\cite{lodi2021one}.

Recent investigations on pattern formation in networks have been restricted to clustering of nodes based on synchronization \cite{pecora2014cluster, sorrentino2016complete, cho2017stablechimera, kumar2024symmetry, bollt2023fractal, Siddique2018symmetry}.  
However, the emergence of patterns such as chimera death states~\cite{poel2015partial, zakharova2014chimera}, which display striking similarities with neuronal states such as unihemispheric sleep in aquatic mammals and birds ~\cite{RATTENBORG2000behavioral}, asymmetric sleep in human patients with sleep anemia~\cite{glaze2016chimerastates}, and more (or less) active brain regions observed in fMRI images~\cite{Gerdes2010brainactivations, Dedreu2019brainactivity}, among the activity patterns cannot be ruled out. Consequently, clusters of synchronized nodes can exhibit active or inactive states.
We call a cluster active if nodes in it have nonzero velocity, and inactive otherwise. We refer to the collection of all such clusters, where some are active and others may be inactive, as an activity pattern. The activeness feature of nodes, when included, generates activity patterns that are distinct from synchrony patterns and have remained largely unexplored. In particular, the structural and dynamical features essential for these patterns to exist remain elusive. 
Depending on the number of nodes participating in inactive cluster(s), these patterns can be classified as partial and complete amplitude (or oscillation) death states \cite{KOSESKA2013oscillation, saxena2012amplitude, ZOU2021quenching}. Partial death states correspond to activity patterns in which only a fraction of the clusters are in the inactive state, while the extreme case in which all clusters become inactive has been referred to as complete amplitude or oscillation death \cite{KOSESKA2013oscillation, saxena2012amplitude, ZOU2021quenching}.

From a dynamical systems perspective, the difficulty in obtaining inactivity for a fraction of nodes while others are active arises due to their connectivity with active nodes. If a neighbor of two active nodes is desired to be inactive, fluctuations from active nodes must cancel out with each other for the inactive node to remain invariant in time. In this work, we demonstrate that simultaneous invariance of active and inactive clusters of synchronized nodes can be achieved using a network structure with permutation symmetries \cite{pecora2014cluster}, provided that the internal dynamics and coupling functions are odd functions with respect to state variables. Perturbations from active clusters can be offset via antisynchronized clusters, allowing some clusters to remain in the inactive state. We generate these antisynchronized clusters from permutation symmetries of quotient networks obtained from different automorphisms of a full network. Although antisynchronization between oscillating elements has been observed earlier in coupled dynamic systems \cite{Prasad2006phase, chowdhury2020effect}, it has not been utilized to achieve inactivity for a fraction of nodes. Unlike synchrony patterns, which exist for all times, some of the activity patterns may be transient, so a method to find their existence for all times is presented.
Finally, we present a unified stability analysis of these patterns in which the stability of identical synchrony, antisynchrony, and inactive clusters is discussed.

The paper is organized as follows. Section \ref{sec:a general setup} introduces coupled dynamical systems that exhibit identical synchronization. Section \ref{sec:permutation symm and} presents the role of permutation symmetries in obtaining identical synchrony between nodes in a network. Section \ref{sec:node dynamics and coupling funct} presents node dynamics and coupling functions that generate coexisting active–inactive clusters of synchronized nodes. The invariance of these activity patterns is also established. Section \ref{sec:existance_of_patt} presents a method to find the existence of activity patterns for all times.
Section \ref{sec:stability analysis} provides a stability analysis in which the stability of identical synchrony, antisynchrony, and inactive clusters is discussed. Section \ref{sec:models} introduces the dynamical systems used for numerical simulations, specifically the Van der Pol and Stuart–Landau oscillators with appropriate coupling functions. Section \ref{sec:results} presents numerical results and the stability analysis of numerically observed patterns, including the switching of clusters between active and inactive states. Finally, in section \ref{sec:conclusions}, we conclude the findings of the paper. 

\section{A general setup of identical oscillators}\label{sec:a general setup}
We consider a network of identical oscillators with pairwise interactions, whose governing equations are represented as
\begin{equation}\label{eq:model}
    \dot{\x}_i=\F(\x_i)+\sigma \sum_{j=1}^N [A]_{ij} \H(\x_i, \x_j), ~~~~i=1,2, \dots, N,
\end{equation}
where $\x_i \in \mathbb{R}^m$ is the position vector of the $i$th oscillator in phase space, $\F:\mathbb{R}^m \to \mathbb{R}^m$ denotes the isolated dynamics, $\H:\mathbb{R}^{m} \times \mathbf{\mathbb{R}^{m}} \to \mathbb{R}^m$ is the coupling function, and $\sigma  \ge 0$ is the overall coupling strength. The matrix $A$ represents the adjacency matrix with entries $[A]_{ij} \in \{0, 1\}$, where $[A]_{ij}=1$ if the nodes $i$ and $j$ are connected, and $0$ otherwise. We consider undirected networks characterized by $[A]_{ij}=[A]_{ji}$.

\section{Permutation symmetries and identical synchrony}\label{sec:permutation symm and}

For identical coupled oscillators in Eq.~\eqref{eq:model}, automorphisms or permutation symmetries of the network structure form the basis of the identically synchronized clusters that a network can exhibit \cite{pecora2014cluster}. An automorphism $\Pi$ of a graph $\mathcal{G}$ is a permutation of all nodes $i \mapsto \Pi(i)$ that preserves the adjacency matrix, i.e., $[A]_{ij}=[A]_{\Pi(i) ~\Pi(j)}$ for all $i,j$. Nodes permuted with each other form identically synchronized clusters. If a network can be split into $k~(1\le k \le N)$ synchronized clusters \cite{sorrentino2016complete}, we denote them as $\mathfrak{C}=\{\mathcal{C}_1, \mathcal{C}_2, \ldots, \mathcal{C}_k\}$. Note that $C_i \cap C_j =\emptyset$ if $i \ne j$ and $\cup_{i = 1}^{k} \mathcal{C}_i=[N]$, where $[N]:=\{1, . . ., N\}$ represents the set of all nodes. The set of all automorphisms under composition forms a group, called $G=\text{Aut}(\mathcal{G})$. Each automorphism can be represented by a permutation matrix $P$, where entries $P_{ij}=1$ if $i \rightarrow j$ during the permutation and $P_{ij}=0$ otherwise. All permutation matrices associated with automorphisms commute with the adjacency matrix, i.e., $AP=PA,~\forall P~\in~G$.
Equation.~\eqref{eq:model} in the vector form is
\begin{equation*}
\dot \x=\tilde{\F}(\x),
\end{equation*}
where 
\begin{align*}
\x&=[\x_1^T, \ldots, \x_N^T]^T \in \mathbb{R}^{Nm},\\
\tilde{\F}(\x)&=[\tilde{\F}_1(\x)^T, \ldots,\tilde{\F}_N(\x)^T]^T \in \mathbb{R}^{Nm},\\
\tilde{\F}_i(\x)&=\F(\x_i)+\sigma \sum_{j=1}^N [A]_{ij} \H(\x_i, \x_j) \in \mathbb{R}^{m}.
\end{align*}
The vector field $\tilde{\F}(\x)$ satisfies the equivariance relation $\tilde{\F}((P \otimes I_m)\x)=(P \otimes I_m)\tilde{\F}(\x)$ for all $P \in G$, where $I_m$ is the identity matrix of size $m \times m$. Each automorphism of the network generates a synchrony pattern which is a fixed point subspace of $P$, defined as
\begin{equation*}
\mathrm{Fix}(P)=\{\x: (P \otimes I_m)\x=\x, ~P~ \in G\}
\end{equation*}
If $(P \otimes I_m)\x(0)=\x(0)$, then the equivariance relation $\tilde{\F}((P \otimes I_m)\x)=(P \otimes I_m)\tilde{\F}(\x)$ guarantees that $(P \otimes I_m)\x(t)=\x(t) ~\forall ~t>0$, i.e. the subspace defined by $\mathrm{Fix}(P)$ is invariant. Note that multiple automorphisms can correspond to the same synchrony pattern. We restrict ourselves to symmetry-based clusters only. However, the coupling functions can also induce clusters ~\cite{schaub2016graph, sorrentino2016complete, Stewart2003symmetry}, which are not covered here.
\subsection*{Pattern dynamics}
In a given synchrony pattern $\mathcal{P}$, the dynamics of the coupled dynamical system, Eq.~\eqref{eq:model}, reduces to that of a quotient network: a smaller network in which each cluster acts as one node. If a network consists of $k$ synchronized clusters, where $\x_i=\s_l~\forall i \in \mathcal{C}_l$, the quotient dynamics can be described by a state vector $\s=[\s_1^T,\dots,\s_k^T]^T \in \mathbb{R}^{km}$. An $i$th node in the quotient network evolves as
\begin{equation}\label{eq:quotient_dyn}
    \dot{\s}_i=\F(\s_i)+\sigma \sum_{j=1}^{k} [Q]_{ij} \H(\s_i, \s_j), ~~~~i=1,2, \dots, k.
\end{equation}
Note that $Q$ represents the quotient matrix with entries $[Q]_{lm}$ given as
\begin{equation*}\label{eq:q_matrix}
    [Q]_{lm}= \frac{1}{|{\mathcal{C}_l}|} \sum_{\substack{i\\ i \in {\mathcal{C}_l}}} \sum_{\substack{j\\ j \in \mathcal{C}_m}} [A]_{ij},
\end{equation*}
where $|.|$ denotes the size of a cluster.

\section{Active-inactive clusters}\label{sec:node dynamics and coupling funct}

Identical synchrony is an invariant state for nodes that have permutation symmetry. However, for general $\F$ and $\H$, nodes in synchronized clusters remain active in time. We call a kth cluster as active if $\dot \x_i  \ne 0 ~\forall ~i \in \mathcal{C}_k$, and inactive otherwise.
For an invariant cluster to be in the inactive state, $\F$ and $\H$ should be selected such that inactivity is an invariant state.
To achieve that, we restrict ourselves to dynamical systems in which $\F$ and $\H$ in Eq.~\eqref{eq:model} satisfy the  criteria
\begin{align}\label{eq:node_and_coupling_funct} 
\begin{split}
    \F(-\x) &= - \F(\x), \\
    \H(-\x_i, -\x_j) &= - \H(\x_i, \x_j),
    \end{split}
\end{align}
which are indeed odd functions in the phase space and imply that $\F(0)=0$ and $\H(0,0)=0$. With these $\F$ and $\H$, the dynamical system \eqref{eq:model} exhibits a steady state solution $\x_i=0 ~\forall ~i \in [N]$, also known as complete amplitude death \cite{ZOU2021quenching}, which does not require permutation symmetries and is an invariant state. Real-world dynamical systems, such as Van der Pol oscillators, Stuart-Landau oscillators \cite{ZOU2021quenching}, identical Kuramoto oscillators with and without inertia \cite{rodrigues2016kuramoto}, small-amplitude oscillations in coupled pendulums and spring-coupled blocks, etc., are some of the paradigmatic models, where individual dynamics and interactions satisfy Eq.~\eqref{eq:node_and_coupling_funct}. 

The patterns of coexisting active-inactive clusters or antisynchronized clusters can be obtained from the graph automorphisms as follows. From each element of the group $G$, we create a quotient network. In this quotient network, the nodes with permutation symmetries but still considered different clusters, i.e., symmetry is broken, may exhibit antisynchronized dynamics in coupled dynamical systems described by Eqs.~\eqref{eq:model} and \eqref{eq:node_and_coupling_funct}. Alternatively, we can find the automorphisms of the quotient network, which form a group $G_q$, and the quotient nodes with permutation symmetries may get antisynchronized. 
To further characterize coexisting active–inactive cluster states on the quotient network, we establish the odd equivariance property of the quotient vector field under automorphisms and show the invariance of activity patterns.

\begin{proposition}[Odd equivariance of the quotient vector field]\label{proposition_equivariance}
Consider the quotient dynamics given by Eq.~\eqref{eq:quotient_dyn} and node dynamics and coupling functions satisfying the odd relation, Eq.~\eqref{eq:node_and_coupling_funct}. Let $P_q$ be a permutation matrix corresponding to an automorphism of the quotient network, i.e., $P_qQ = QP_q, ~P_q \in G_q$. Define the stacked quotient vector field
\begin{align*}
\tilde{\F}_q(\s)&=[[{\tilde{\F}_q}(\s)]_1^T,\dots,[{\tilde{\F}_q}(\s)]_k^T]^T \in \mathbb{R}^{km},\\
[{\tilde{\F}_q}(\s)]_i&=\F(\s_i)+\sigma \sum_{j=1}^{k} [Q]_{ij} \H(\s_i, \s_j) \in \mathbb{R}^{m}.
\end{align*}
Then the vector field $\tilde \F_q(\s)$ satisfies the equivariance relation
\begin{equation*}
\tilde{\F}_q(-(P_q\otimes I_m)\s)=-(P_q\otimes I_m)\tilde{\F}_q(\s).
\end{equation*}
\end{proposition}

\begin{proof}
The $i$-th block of $\tilde{\F}_q(-(P_q\otimes I_m)\s)$ is
\[
[\tilde{\F}_q(-(P_q\otimes I_m)\s)]_i={\F}(-\s_{\pi(i)})+\sigma\sum_{j=1}^k [Q]_{ij}\H(-\s_{\pi(i)},-\s_{\pi(j)}),
\]
where $\pi$ is the permutation associated with $P_q$. Using the odd symmetry of $\F$ and $\H$ from Eq.~\eqref{eq:node_and_coupling_funct}, we obtain
\[
[\tilde{\F}_q(-(P_q\otimes I_m)\s)]_i=- {\F}(\s_{\pi(i)})-\sigma\sum_{j=1}^k [Q]_{ij}\H(\s_{\pi(i)},\s_{\pi(j)}).
\]
Since $P_q$ corresponds to the quotient automorphism $\pi$ and $P_qQ=QP_q$, we have $[Q]_{ij}=[Q]_{\pi(i)\pi(j)}$, which yields
\[
[\tilde{\F}_q(-(P_q\otimes I_m)\s)]_i=-[\tilde{\F}_q(\s)]_{\pi(i)},
\]
The RHS is $[-(P_q\otimes I_m)\tilde{\F}_q(\s)]_i$. Since this holds for all $i$, we get
\[
\tilde{\F}_q(-(P_q\otimes I_m)\s)=-(P_q\otimes I_m)\tilde{\F}_q(\s).
\]
\end{proof}
\begin{figure*}
 \includegraphics[width=0.9\linewidth]{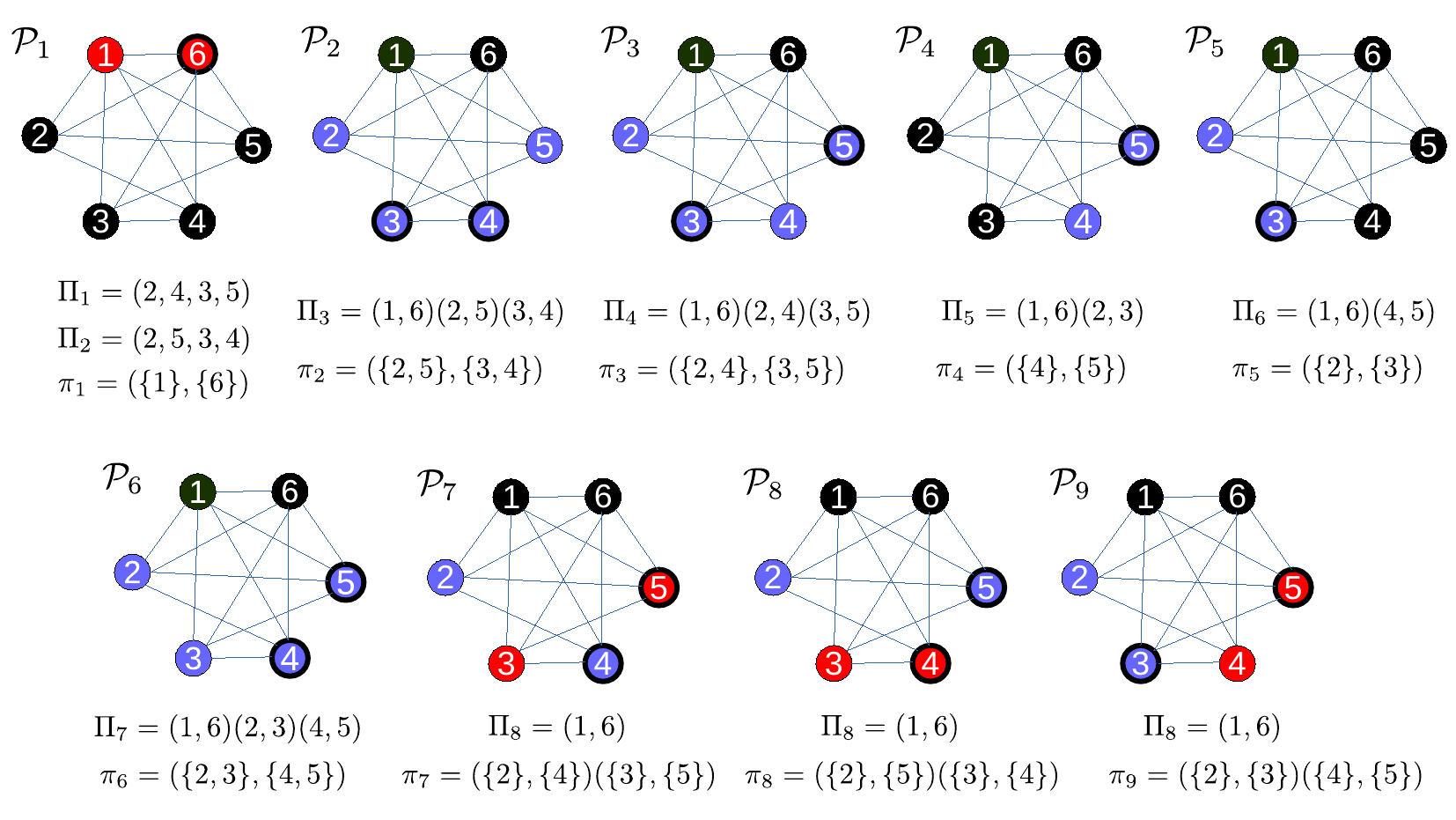}
 \caption{Invariant patterns containing active-inactive clusters in a network of $6$ nodes. 
 Synchronized nodes are shown in the same color and boundary thickness. Clusters shown in black correspond to the amplitude death state, while those in blue or red can be either active or in oscillation death state. If two clusters are antisynchronized with each other, they are shown in the same color, but one of them has a thicker boundary line. The full network automorphism ($\Pi$) and the quotient automorphism ($\pi$) that generate a pattern are shown below each pattern. Nodes permuted with themselves are not shown for brevity. The quotient nodes are clusters in general, so they are denoted as sets, regardless of whether a cluster is trivial or nontrivial.}
 \label{fig:activity_patterns}
 \end{figure*}
\begin{proposition}[Invariance of activity patterns]
Let $P_q$ be a permutation matrix corresponding to an automorphism of the quotient network. Define
\begin{equation}\label{eq:manifold_fix_p_q}
\mathrm{Fix}(P_q)=\{\s: (P_q \otimes I_m)\s=-\s, ~    P_q \in G_q\}.
\end{equation}
If the initial condition satisfies $(P_q \otimes I_m)\s(0)=-\s(0)$ then $(P_q \otimes I_m)\s(t)=-\s(t)~\forall t>0$, i.e., the manifold $Fix(P_q)$ is invariant under the quotient dynamics (Eq.~\eqref{eq:quotient_dyn}). 
\end{proposition}
\begin{proof}
Since the manifold is given by $(P_q \otimes I_m) \s=-\s$, the equivariance relation from proposition \ref{proposition_equivariance}
\[
\tilde{\F}_q(-(P_q\otimes I_m)\s)=-(P_q\otimes I_m)\tilde{\F}_q(\s)
\]
yields
\[
\tilde{\F}_q(\s)=-(P_q\otimes I_m)\tilde{\F}_q(\s).
\]
Since $\dot \s=\tilde\F_q(\s)$, we obtain
\[
(P_q \otimes I_m) \dot \s=-\dot \s,
\]
which shows that the velocity field satisfies the same constraint as the synchrony subspace $\s$. Hence, the manifold $\mathrm{Fix}(P_q)$ is invariant.
\end{proof}

\begin{remark*}[Coexisting active-inactive clusters]
We look for quotient automorphisms that can generate antisynchrony between the nodes. Let $\pi$ be a quotient automorphism whose cycle decomposition consists only of
1-cycles and 2-cycles, i.e.,
\[
\pi=(i_1,j_1)(i_2,j_2), \ldots, (i_r,j_r)(l_1)(l_2), \ldots,(l_s),
\]
where $2r+s=k$. The automorphism $\pi$ either exchanges pairs of clusters or leaves clusters fixed. If $P_q$ is the permutation matrix corresponding to the automorphism $\pi$, consider the manifold $\mathrm{Fix}(P_q)$ given in Eq.~\eqref{eq:manifold_fix_p_q}, which shows that clusters permuted with each other by $P_q$ form antisynchronized clusters satisfying $\s_i=-\s_j$, while those mapped to themselves satisfy $\s_i=0$ and correspond to the inactive clusters. The matrices $P_q$ satisfy $P_q^2=I_k$, i.e., they are self-inverse (involutory).
\end{remark*}

Below, we show activity patterns of coexisting active-inactive clusters or antisynchronized clusters for the network shown in Fig.~\ref{fig:activity_patterns}. We find the automorphisms of the full network and generate quotient networks corresponding to each automorphism, so synchronization between network nodes is taken care of by full network automorphisms. Involutory automorphisms of the quotient networks generate patterns of coexisting active-inactive clusters. It can be shown that the subspace $(P_q \otimes I_m) \s=\s$ is also invariant, so involutory automorphisms also correspond to synchronization between quotient nodes, but we only use them to find antisynchrony between quotient nodes.
The network in Fig.~\ref{fig:activity_patterns} has $16$ automorphisms, which can be computed using openly available discrete algebra tools \cite{pecora2014cluster, GAP2005, Stein2013Sage}. The automorphism group $G$ in terms of its generators is 
\[
G=<(2,3), (4,5), (2,4)(3,5), (1,6)>. 
\]
Pattern $\mathcal{P}_{1}$ in Fig.~\ref{fig:activity_patterns} corresponds to automorphisms $(2,4,3,5)$ and $(2,5,3,4)$, and the quotient automorphism $(\{1\},\{6\})$ generate antisynchrony between clusters $\{1\}$ and $\{6\}$. The cluster $\{2,3,4,5\}$ permutes with itself and so forms the inactive state. In general, the quotient nodes are clusters, so we denote them as clusters, whether a node is a trivial cluster or nontrivial. Pattern $\mathcal{P}_2$ corresponds to network automorphism $(1,6)(2,5)(3,4)$ and the quotient automorphism $(\{2,5\})(\{3,4\})$ generates antisynchrony between the permuting clusters. An additional eight automorphisms, two of which are analogous to $\Pi_1$ and $\Pi_2$, but also permute nodes $1$ and $6$, lead to complete amplitude death states since the corresponding quotient nodes can swap with themselves only. Five automorphisms, analogous to $\Pi_3-\Pi_7$ but nodes $1$ and $6$ permute with themselves, generate patterns similar to $\mathcal{P}_2-\mathcal{P}_6$ but the quotient nodes $\{1\}$ and $\{6\}$ exhibit antisynchronization. Lastly, the identity automorphism, in which each network node permutes with itself, generates patterns similar to $\mathcal{P}_7-\mathcal{P}_9$, but clusters $\{1\}$ and $\{6\}$ exist in antisynchrony. Note that multiple automorphisms of the quotient networks may generate the same activity pattern. For instance, patterns $\mathcal{P}_4$ and $\mathcal{P}_5$ can be generated from the quotient network corresponding to $\Pi_8$. Further note that, although structurally different, multiple clusters can exhibit the same state (amplitude death state), such as clusters $\{1,6\}$ and $\{2,3\}$ in pattern $\mathcal{P}_4$. 

In Fig.~\ref{fig:activity_patterns}, inactive clusters in amplitude death state are shown in black, while those in blue or red can be either active or inactive (oscillation death states, i.e., $\dot \x_j =0$ but $\x_j \ne 0$ $~\forall ~j \in \mathcal{C}_i$, provided that the oscillation death solutions exist). 
A key property of the activity patterns is that each active cluster in a network must be accompanied by an antisynchronized cluster, or mathematically, $\sum_{i=1}^N\dot  \x_i=0$, i.e., total velocity is always conserved. Despite the oscillatory nature of some clusters, the inactivity a cluster is possible due to the antisynchrony between active clusters, which results in the null value being input to their common neighbor, facilitating their neighbor to remain in the inactive state.

All the patterns of coexisting active-inactive clusters can also be obtained from orbital clusters. Since antisynchronized clusters have permutation symmetries, we can find them from the symmetry breaking of orbital clusters. The orbit of a node $i$ is defined as the set of nodes that node $i$ can be mapped to under the action of all automorphisms, i.e., $\mathrm{Orb}(i)=\{\pi(i)~ : \pi~\in~G \}$, resulting in the orbital cluster containing node $i$. 
Thereafter, the identical synchrony between nodes in the orbital clusters is broken, and antisynchronized clusters are created such that invariant clusters connected with these antisynchronous clusters can remain inactive. The process is repeated until all possible patterns are obtained. For example, patterns $\mathcal{P}_{1}$ and $\mathcal{P}_2-\mathcal{P}_9$ are generated from the symmetry breaking of the orbital clusters $\{1,6\}$ and $\{2,3,4,5\}$.

\section{Existence of patterns}\label{sec:existance_of_patt}

Except for the complete amplitude-death state, the invariant patterns of coexisting active-inactive clusters, or clusters antisynchronized with one another, may be transient because clusters in the oscillatory state may transit to the amplitude or oscillation-death state as time progresses. 
Starting with an activity pattern $\mathcal{P}_i$, if a pair of antisynchronized clusters can not sustain their oscillations and transitions to the amplitude or oscillation-death state (corresponding to pattern $\mathcal{P}_j$), the pattern $\mathcal{P}_j$ exists in the network. For instance, starting with pattern $\mathcal{P}=\{\x_1, \x_2,-\x_2,\x_4,-\x_4,-\x_1\}$, if clusters $\{1\}$ and $\{6\}$ transitions to the amplitude death state at a given $\sigma$, pattern $\mathcal{P}_3$ exist at that $\sigma$ value. 

In the dynamical system described by Eqs.~\eqref{eq:model} and \eqref{eq:quotient_dyn}, we take the pattern $\mathcal{P}_i$ with corresponding $\s=[\s_1^T, \ldots, \s_l^T,\s_m^T, \ldots, \s_k^T]^T$ at a coupling $\sigma$. If the antisynchronized pair $\s_l$ and $\s_m$, where $\s_l=-\s_m$, transitions to an inactive state $\s_l^*=-\s_m^*$ corresponding to pattern $\mathcal{P}_j$, we perturb the state vector $\s'$ corresponding to $\mathcal{P}_j$ as $\s' \rightarrow \s'+\delta\s$, where 
\[\s'=[\s_1^T, \ldots, (\s_l^*)^T,(-\s_l^*)^T, \ldots, \s_k^T]^T
\]
and 
\[\delta\s=[0, \ldots, \delta\s_l^T,-\delta\s_l^T, \ldots, 0]^T.\] 
If $\delta\s \rightarrow 0$ as $t \rightarrow \infty$, pattern $\mathcal{P}_j$ exist at coupling $\sigma$. Note that if the transition occurs to the amplitude death state, $\s_l^*=-\s_m^*=0$.

In general, a cluster dynamics in a pattern is coupled with other clusters. However, if an active cluster $\mathcal{C}_i$ receives net-zero input from other clusters (except its antisynchronized counterpart, cluster $\mathcal{C}_j$) and the coupling function can be written as $\H(\s_i, \s_j)=\H^1(\s_i)+\H^2(\s_j)$, the dynamics of such an oscillating cluster (and its antisynchronized counterpart) is independent of the full quotient dynamics. For numerical simulations, we restrict ourselves to such networks only. From Eq.~\eqref{eq:quotient_dyn}, the dynamics of the cluster $\mathcal{C}_i$ in an invariant state is given as
\begin{align}\label{eq:indep_clus_dyn_1}
    \dot{\s}_i=\F(\s_i)+\sigma \Bigl\{ [Q]_{ii}(\H^1(\s_i)+\H^2(\s_i)) \nonumber\\+[Q]_{ij}(\H^1(\s_i)+\H^2(\s_j))  +\sum_{l=1, l \ne i,j}^{k} [Q]_{il} (\H^1(\s_i)+\H^2(\s_l))\Bigr\}.
\end{align}
Since $\s_j=-\s_i$ and $\sum_{l=1, l \ne i,j}^{k} [Q]_{il} \H^2(\s_l)=0$ because of the assumption, cluster $\mathcal{C}_i$ evolve independently of the other quotient nodes. 
In such a case, if the antisynchronized pair $\mathcal{C}_i$ and $\mathcal{C}_j$ transition to an inactive state (amplitude or oscillation death state), Eq.~\eqref{eq:indep_clus_dyn_1} satisfies
\begin{align}\label{eq:indep_clus_dyn_2}
    \dot{\s}_i=0=\F(\s_i)+\sigma \Bigl\{ [Q]_{ii}(\H^1(\s_i)+\H^2(\s_i))\nonumber\\+[Q]_{ij}(\H^1(\s_i)+\H^2(-\s_i))+\sum_{l=1, l \ne i,j}^{k} [Q]_{il} \H^1(\s_i)\Bigr\}.
\end{align}
The stability of the solution (corresponding to amplitude or oscillation death state) in Eq.~\eqref{eq:indep_clus_dyn_2} determines if oscillations can be sustained in clusters $\mathcal{C}_i$ and $\mathcal{C}_j$. The collective state of all clusters in the network determines the existence of a pattern.

\section{Stability analysis}\label{sec:stability analysis}

To determine the stability of an invariant pattern $\mathcal{P}$, we perturb
$\x \in \mathcal{P}$ such that $\x \rightarrow \x +\deltax$, where $\deltax =[\deltax_1^T, \ldots, \deltax_N^T]^T \in \mathbb{R}^{Nm}$. From Eq.~\eqref{eq:model}, the evolution equation of the perturbation vector $\deltax$ can be obtained as
\begin{align}\label{eq:perturb_dyn}
    \dot{\deltax}&=\biggl\{ \sum_{l=1}^{k} E^l \otimes D\F(\s_l) \nonumber \\ 
    &+\sigma \sum_{l=1}^{k} \sum_{m=1}^{k} E^l A E^m \otimes D\H^1(\s_l, \s_m) \nonumber \\ 
    &+ \sigma \sum_{l=1}^{k} E^l \otimes \sum_{m=1}^{k} Q_{lm} D\H^2(\s_l, \s_m) \biggl\} \deltax, 
\end{align}
where we have neglected higher-order terms in the Taylor expansion of Eq.~\eqref{eq:model} and
\begin{align*}
    D\F(\s_i)=\left. \frac{\partial \F(\x)}{\partial \x} \right \vert_{\s_i}, && D\H^{1(2)}(\s_i,\s_j)=\left. \frac{\partial \H(\x_i, \x_j)}{\partial \x_{j(i)}} \right \vert_{\s_i, \s_j}.\\
\end{align*}
The diagonal matrices $E^j, j \in [k]$, are defined such that
\begin{align*}
[E^j]_{ii}=\begin{cases}
    1 & \text{if} ~~~~ \text{i} \in \mathcal{C}_j,\\
    0 & \text{otherwise}.
\end{cases}
\end{align*}
To determine the stability of an invariant cluster, we must decouple transversal perturbations from the cluster synchronization manifold. For this purpose, we make a coordinate transformation such that $\deltax = (T^{-1} \otimes I_m) \deltaeta$, where $\deltaeta=[\deltaeta_1^T, \ldots, \deltaeta_N^T]^T \in \mathbb{R}^{Nm}$. The first $k$ rows in $T \in \mathbb{R}^{N \times N}$ represents modes longitudinal to synchronization subspace, while the last $N-k$ rows corresponds to transversal modes which breaks identical synchrony between nodes, so $T$ can be broken as $T = [T_{||}^T ~T_{\perp}^T]^T$, where elements of an $i$th row in $T_{||}$ are such that $[{T_{||}}]_{ij}=1/\sqrt{|\mathcal{C}_i|}$ if $j \in \mathcal{C}_i$ and zero otherwise. The part $T_{\perp}$ is obtained using a method of symmetry breaking of invariant clusters \cite{lodi2021one}. Other methods, such as irreducible representation of symmetry groups \cite{pecora2014cluster}, and simultaneous block diagonalization of the coupling matrix \cite{Zhang2020symmetry} have also been used to achieve the block diagonalization of the coupling matrix. 
Using this transformation, Eq.~\eqref{eq:perturb_dyn} can be expressed as
\begin{align}\label{eq:perturb_dyn_eta_coordinates}
    {\dot{\deltaeta}}&=\biggl\{ \sum_{l=1}^{k} (E^l)^* \otimes D\F(\s_l) \nonumber \\
    &+\sigma \sum_{l,m=1}^{k}  (E^l A E^m)^* \otimes D\H^1(\s_l, \s_m) \nonumber \\ 
    &+ \sigma \sum_{l,m=1}^{k} (E^l)^* \otimes  Q_{lm} D\H^2(\s_l, \s_m) \biggl\} \deltaeta,
\end{align}
where $(.)^*= T(.)T^{-1}$. If we denote the bracket term in RHS in Eq.~\eqref{eq:perturb_dyn_eta_coordinates} by $C$, it has the form \cite{lodi2021one}
\begin{align}\label{eq:block_B}
C = \begin{bmatrix}
    C^{\parallel} & 0\\ 
    0 & C^{\perp}  \\
    \end{bmatrix},
\end{align}
where $C^{\parallel} \in \mathbb{R}^{km \times km}$ and $C^{\perp} \in \mathbb{R}^{(N -k)m \times (N -k)m}$. 

\subsection{Stability of identical synchronization}
The matrix $C^{\perp}$ is block diagonal and each block corresponds to a cluster or set of clusters if they are intertwined \cite{lodi2021one, della2020symmetries, pecora2014cluster}. A cluster $\mathcal{C}_k$ of size $|\mathcal{C}_k|$ has one direction ($\deltaeta_k$) parallel to the synchronization manifold and $|\mathcal{C}_k|-1$ transverse directions in the node space ($\deltaeta_k^{i}, ~i=1,2, \ldots |{\mathcal{C}_k}|-1$) that determine the stability of identical synchrony between nodes. Therefore, if all $\deltaeta_k^{i} \rightarrow 0~$ as $t \rightarrow \infty$, identical synchrony between nodes in cluster $\mathcal{C}_k$ is stable. 

\subsection{Stability of inactive clusters and antisynchronization}\label{subsec:stability inactive clusters}
The stability of fixed point states (or inactive clusters) and antisynchrony between clusters cannot be determined from ${C}^{\perp}$. For this purpose, we resort to perturbations in ${C}^{\parallel}$, which contain $k$ directions parallel to each cluster manifold. If the identical synchrony between nodes in a cluster $\mathcal{C}_l$ is stable, then  all $\deltaeta_l^{i}$ associated with the cluster $\mathcal{C}_l$ decay with time, and therefore, $\deltax_i \rightarrow \deltax_l ~\forall~i \in \mathcal{C}_l$. Now, for this cluster to be in a fixed point state, we require $\deltaeta_l \propto \deltax_l \rightarrow 0$ as $t \rightarrow \infty$, which we find from solving $\deltaeta_l$. To determine the stability of antisynchrony between clusters $\mathcal{C}_l$ and $\mathcal{C}_m$, we solve $\deltaeta_l+\deltaeta_m$. If $\deltaeta_l+\deltaeta_m \propto \deltax_l+\deltax_m \rightarrow 0$ as $t \rightarrow \infty$, the antisynchrony between the clusters $\mathcal{C}_l$ and $\mathcal{C}_m$ is also stable. 

We can write $\deltaeta=[\deltaeta_{||}^T ~\deltaeta_{\perp}^T]^T$ and the modes determining stability of inactive clusters and antisynchrony in $\deltaeta_{||}$ can be decoupled from unwanted modes as follows. The longitudinal modes in Eq.~\eqref{eq:perturb_dyn_eta_coordinates} evolve as
\begin{align}\label{eq:perturb_dyn_long_modes}
    {\dot{\deltaeta}_{||}}&=\biggl\{ \sum_{l=1}^{k} (E^l)^* \otimes D\F(\s_l) \nonumber \\
    &+\sigma \sum_{l,m=1}^{k}  (E^l A E^m)^* \otimes D\H^1(\s_l, \s_m) \nonumber \\ 
    &+ \sigma \sum_{l,m=1}^{k} (E^l)^* \otimes  Q_{lm} D\H^2(\s_l, \s_m) \biggl\} \deltaeta_{||},
\end{align}
where $(.)^*= T_{||}(.)T_{||}^{T}$. Eq.~\eqref{eq:perturb_dyn_long_modes} has an additional symmetry which we show as follows: First, note that, for any $l$, $T_{||}E^lT_{||}^{T}=\operatorname{diag}(d_1, \ldots,d_m,\ldots, d_k) \in \mathbb{R}^{k \times k}$, where $d_m=1$ if $m=l$ and zero otherwise. Now,
\begin{align*}
T_{||}E^lAE^mT_{||}^{T}=(T_{||}E^lT_{||}^{T})T_{||}AT_{||}^{T}(T_{||}E^mT_{||}^{T})
\end{align*}
We can write 
\begin{align*}
T_{||}AT_{||}^{T}=M \tilde{T_{||}}A(M\tilde{T_{||}})^{T},
\end{align*}
where $T_{||}=M\tilde{T_{||}}$ and $M =\operatorname{diag}(1/\sqrt{|\mathcal{C}_1}|, \ldots, 1/\sqrt{|\mathcal{C}_k}|) \in \mathbb{R}^{k \times k}$. The matrices $A, Q$ and $\tilde{T_{||}}$ are related as $A(\tilde T_{||})^T=(\tilde T_{||})^TQ$. Using this relation,
\begin{align*}
T_{||}AT_{||}^{T}=M\tilde T_{||}(\tilde T_{||})^TQM^{T}=M^{-1}QM,
\end{align*}
which shows $T_{||}AT_{||}^{T}$ and the quotient matrix $Q$ are similar matrices. For a symmetry based inactive cluster $\mathcal{C}_k$ and antisynchronized clusters $\mathcal{C}_l$ and $\mathcal{C}_m$, we have $[T_{||}AT_{||}^{T}]_{kl}=[T_{||}AT_{||}^{T}]_{km}$, $[T_{||}AT_{||}^{T}]_{lk}=[T_{||}AT_{||}^{T}]_{mk}$, $[T_{||}AT_{||}^{T}]_{lm}=[T_{||}AT_{||}^{T}]_{ml}$, and $[T_{||}AT_{||}^{T}]_{ll}=[T_{||}AT_{||}^{T}]_{mm}$.  Furthermore, 
\begin{equation*}\label{eq:df_and_dh_symmetry}
\begin{split}
D\F(-\s)&=D\F (\s),\\
D\H^{1(2)} (-\s_i, -\s_j)&=D\H^{1(2)} (\s_i, \s_j), 
\end{split}
\end{equation*}
Using these structural and dynamical symmetries, the transversal mode $\deltaeta_l +\deltaeta_m$ and longitudinal mode(s) associated with inactive cluster(s) can be decoupled from $\deltaeta_l -\deltaeta_m$, the mode longitudinal to the antisynchronization subspace and not required for the stability of pattern $\mathcal{P}$. For this purpose, we make a coordinate transformation such that $\deltaeta \rightarrow (S_R^{-1} \otimes I_m) \deltaxi$, where $\deltaxi =[\deltaxi_1^T, \ldots, \deltaxi_{(N-n_a)}^T]^T \in \mathbb{R}^{(N-n_a)m}$. The matrix $S_R^{-1}$ is the right inverse of a matrix $S\in \mathbb{R}^{(N-n_a) \times N}$, where $n_a$ denotes number of antisynchronized pairs of clusters. The matrix $S$ is obtained from an identity matrix $I_N$ of size $N \times N$ such that, if $\deltaeta_l$ and $\deltaeta_m$ are summed, its $l$th row is given by $[S]_{lj}=[I_N]_{lj}+[I_N]_{mj}$, where $j \in [N]$. All other rows of $S$ are the same as those of $I_N$ except that the $m$th row of $I_N$ should be deleted. Finally, the row vectors in $S$ are normalized to unity 2-norm. Note that $l$ and $m$ are interchangeable. The matrix $S$ is real and has orthogonal rows, and therefore, there exists a right inverse such that $SS_R^{-1}=I_{N-n_a}$, where $S_R^{-1}=S^T(SS^T)^{-1} \in \mathbb{R}^{N \times (N-n_a)}$. Using this transformation, Eq.~\eqref{eq:perturb_dyn_eta_coordinates} can be written as
\begin{align}\label{eq:perturb_dyn_xi_coordinates}
    {\dot{\deltaxi}}&=\biggl\{ \sum_{l=1}^{k} (E^l)^{'} \otimes D\F(\s_l) \nonumber \\
 &+\sigma  \sum_{l,m=1}^{k} (E^l A E^m)^{'} \otimes D\H^1(\s_l, \s_m) \nonumber \\ 
  &  + \sigma \sum_{l,m=1}^{k} (E^l)^{'} \otimes Q_{lm} D\H^2(\s_l, \s_m) \biggl\} \deltaxi,
\end{align}
%
\begin{figure*}
 \includegraphics[width=\textwidth]{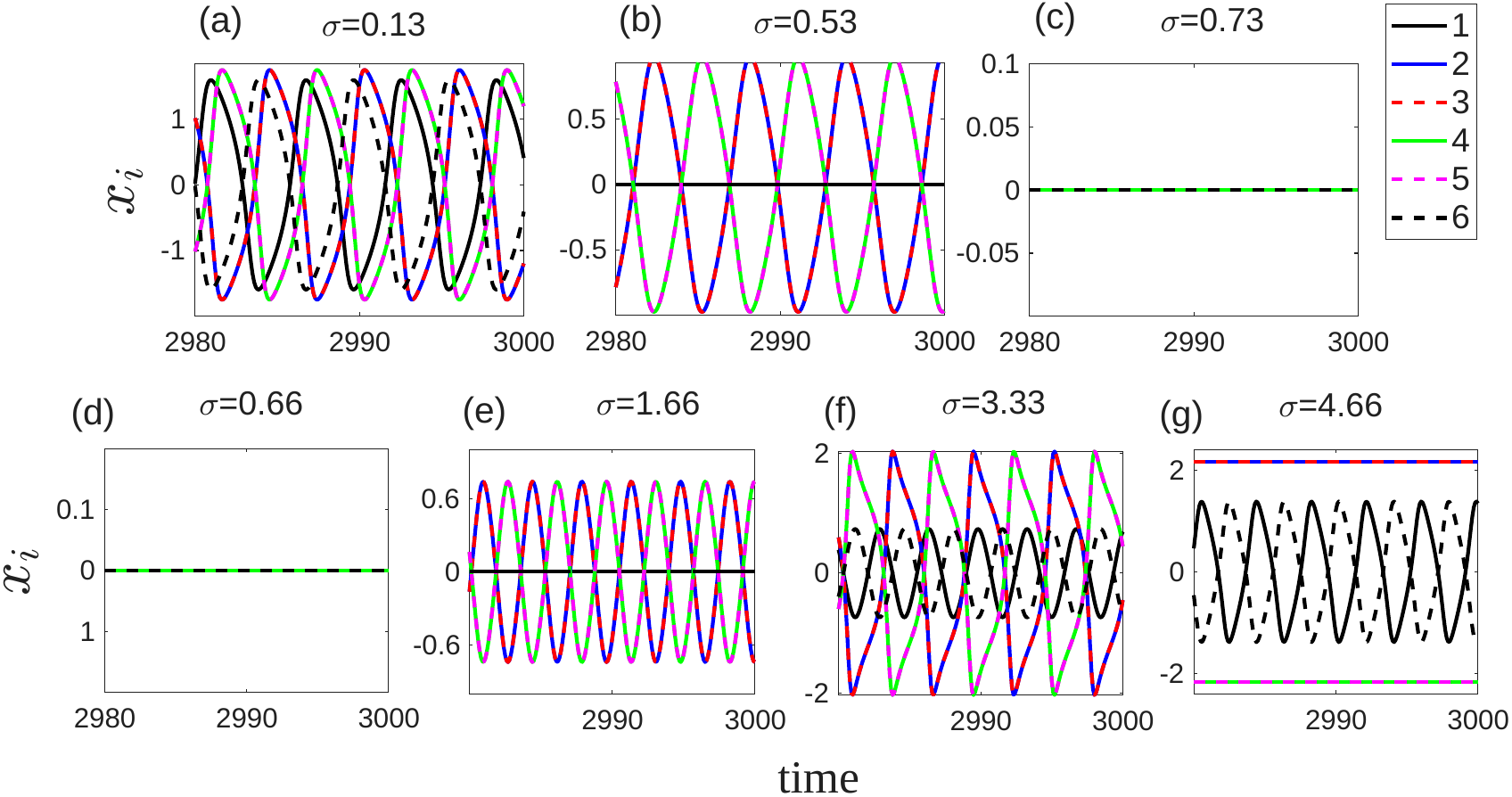}
 \caption{Time series shows activity patterns for the 6-node network presented in Fig.~\ref{fig:activity_patterns}. (a)-(c) The network of Van der Pol oscillators switches between patterns $\mathcal{P}_{10} \rightarrow \mathcal{P}_6 \rightarrow \mathcal{P}_{0}$ when $\sigma$ is increased. (d)-(g) The network of Stuart-Landau oscillators switches between patterns $\mathcal{P}_{0} \rightarrow \mathcal{P}_6 \rightarrow \mathcal{P}_{10} \rightarrow \mathcal{P}_{11}$ as a function of $\sigma$.}
 \label{fig:time_series}
 \end{figure*}

where $(.)^{'}= ST(.)T^{-1} S_R^{-1}$. The bracket term in the RHS of Eq.~\eqref{eq:perturb_dyn_xi_coordinates} still has the same form as in Eq.~\eqref{eq:block_B} but $C^{\parallel} \in \mathbb{R}^{(k-n_a)m \times (k-n_a)m}$.

Therefore, $C^{\perp}$ together with $C^{\parallel}$ provides complete insight into the stability of a pattern consisting of active-inactive clusters. A single matrix $T'=ST$ is sufficient to determine the stability of all three synchrony arrangements, i.e., synchrony, antisynchrony, and inactive state of clusters that the network can exhibit. For example, in pattern $\mathcal{P}_6$, the $T'$ matrix has the form
\begin{align*}
T'= \frac{1}{\sqrt{4}}
\begin{bmatrix}
\sqrt{2} & 0 & 0 & 0 & 0 & \sqrt{2}\\
0 & 1 & 1 & 1 & 1 & 0\\
-\sqrt{2} & 0 & 0 & 0 & 0 & \sqrt{2}\\
0 & -\sqrt{2} & \sqrt{2} & 0 & 0 & 0\\
0 & 0 & 0 & -\sqrt{2} & \sqrt{2} & 0
\end{bmatrix}.
\end{align*}
We can write $\deltaxi =\{T' \otimes I_m\} \deltax$, so  the identical synchrony between nodes in clusters $\{1,6\}$, $\{2,3\}$ and $\{4,5\}$ is stable if $\deltaxi_3, \deltaxi_4,$ and $\deltaxi_5 \rightarrow 0$, respectively. Furthermore, we require $\deltaxi_1 \rightarrow 0$ for the inactive state to  be stable for $\{1,6\}$, while $\deltaxi_2 \rightarrow 0$ if antisynchronization is stable between $\{2,3\}$ and $\{4,5\}$. 

If a network is in the complete amplitude or oscillation death state, all perturbations $\deltax_i$ must decay with time, and the coordinate transformation $\deltax \rightarrow \deltaxi$ is not required. From a closer look into the RHS in Eq.~\eqref{eq:perturb_dyn_xi_coordinates}, one can check that the stability of inactivity and antisynchrony are intertwined, i.e., the loss of stability of antisynchrony between active clusters destabilizes inactive clusters and vice-versa. Eq.~\eqref{eq:perturb_dyn_xi_coordinates} together with Eq.~\eqref{eq:quotient_dyn} is solved numerically to determine the stability of the pattern $\mathcal{P}$. We track all $(N-n_a)m$ Lyapunov exponents and the largest of all, which we call $\Gamma (\sigma)$, is plotted as a function of $\sigma$. A negative value of $\Gamma (\sigma)$ shows the stability of the invariant pattern $\mathcal{P}$. 

\section{Models}\label{sec:models}

We validate our theory using two paradigmatic models that satisfy Eq.~\eqref{eq:node_and_coupling_funct}. As the first system, we consider  a network of Van der Pol oscillators, where the isolated node dynamics and the coupling function are represented as
\begin{equation}\label{eq:vanderpol_osc}
\F(\x_i)=\begin{bmatrix}
    y_i\\
    \mu(1-x_i^2)y_i-x_i
\end{bmatrix},
\H(\x_i, \x_j)=\begin{bmatrix}
    \sigma_x (x_j -x_i)\\
    \sigma_y (y_j+y_i)
\end{bmatrix},
\end{equation}
where $\mu, \sigma_{x(y)} \in \mathbb{R}$. We choose  $\mu=2$ and $\sigma_{x(y)}=0.2(-1)$.  We consider Stuart-Landau oscillators \cite{ZOU2021quenching} as our second system to corroborate our results, whose node dynamics and the coupling functions are represented as
\begin{equation}\label{eq:sl_osc}
\F(\x_i)=\begin{bmatrix}
    [\lambda -(x_i^2+y_i^2)]x_i -\omega y_i \\
    [\lambda -(x_i^2+y_i^2)]y_i +\omega x_i
\end{bmatrix},
\H(\x_i, \x_j)=\begin{bmatrix}
    \sigma_x x_j\\
    \sigma_y y_j
\end{bmatrix},
\end{equation}
where $\lambda, \omega \in \mathbb{R}$ determine the uncoupled dynamics.  We fix $\lambda=-2, \omega=2$ along with $\sigma_{x(y)}=-1(-0.5)$. The coupling functions $\H(\x_i, \x_j)$ for both systems are selected in these particular forms to ensure that the desired patterns are stable. 

\section{Results}\label{sec:results}

In this section, we show that dynamical systems whose network structure contains permutation symmetries and node dynamics satisfy Eq.~\eqref{eq:node_and_coupling_funct} can display stable activity patterns consisting of both active and inactive clusters of synchronized nodes. As the coupling strength $\sigma$ increases, the active clusters manifest as inactive clusters at different values of $\sigma$. Consequently, the network exhibits different activity patterns as a function of the coupling strength, eventually resulting in a complete amplitude (or oscillation) death state. 

We consider a network of 6 nodes as shown in Fig.~\ref{fig:activity_patterns}.  We solve  Eq.~\eqref{eq:model} using the Runge-Kutta 4th-order method with time-varying integration step sizes. Initial conditions are chosen close to the pattern $\mathcal{P}_{10}=\{(\x_1 ~\x_2 ~\x_2 ~-\x_2 ~-\x_2 ~-\x_1)\}$. The temporal evolution of $x_i$  corresponding to the Van der Pol oscillators is depicted in Figs.~\ref{fig:time_series}(a)-\ref{fig:time_series}(c), while that for the  Stuart-Landau oscillators is depicted in Figs.~\ref{fig:time_series}(d)-\ref{fig:time_series}(g).  Note that $x_i$ are represented using different line styles and colors as shown in the figure legend. 
The network of Van der Pol oscillators initially exhibits the activity pattern $\mathcal{P}_{10}$ (see Fig.~\ref{fig:time_series}(a) for $\sigma=0.13$), where all clusters consist of active nodes.
Other patterns different from  $\mathcal{P}_{10}$ may also exist depending on the specific initial conditions. 
Increasing the coupling strength leads to the manifestation of the pattern $\mathcal{P}_6$, shown in Fig.~\ref{fig:time_series}(b) for $\sigma=0.53$, and subsequently to $\mathcal{P}_{0}$ (complete amplitude death state), depicted in Fig.~\ref{fig:time_series}(c) for $\sigma=0.73$.
Similarly, the network of Stuart-Landau oscillators exhibits a transition from $\mathcal{P}_{0} \rightarrow \mathcal{P}_6 \rightarrow \mathcal{P}_{10} \rightarrow \mathcal{P}_{11}$ as a function of $\sigma$ (see Figs.~\ref{fig:time_series}(d)-\ref{fig:time_series}(g)). 
Inactive clusters in Figs.~\ref{fig:time_series}(b)-\ref{fig:time_series}(e) are in the amplitude death state, while in Fig.~\ref{fig:time_series}(g), they are characterized by the oscillation death state. 
It is evident from  Figs.~\ref{fig:time_series}(b), \ref{fig:time_series}(e), and \ref{fig:time_series}(g) that the synchronized active clusters are counter-balanced by the presence of the anti-synchronized clusters, which leads to cancellation of fluctuations for inactive clusters to coexist.
Note that these results hold for any valid pattern, irrespective of network size. 

Next, we calculate the range of coupling strength, wherein the patterns observed in Fig.~\ref{fig:time_series} are stable. Note that the calculation of the largest Lyapunov exponent $\Gamma(\sigma)$ for a particular invariant pattern necessitates the existence of the corresponding quotient dynamics in \eqref{eq:quotient_dyn}. However, we find that a particular quotient dynamics exists only in a certain range of the coupling strength, except when the network is in complete amplitude death (pattern $\mathcal{P}_{0}$). Therefore, first we find the range of the coupling strength in which different invariant patterns and their corresponding quotient dynamics exist.

\subsection{Transition from activity to inactivity}\label{subsec:transition_from_activity}

Active clusters in the full node space (Eq.~\eqref{eq:model}), as well as in the invariant states (Eq.~\eqref{eq:quotient_dyn}),
manifest as inactive clusters due to the onset of amplitude or oscillation death state as a function of the coupling strength. Consequently, the network exhibits transitions from one activity pattern to another. The coalescence of clusters at $[0~ 0]^T$ is similar to the phenomenon of cluster merging, in which two or more active clusters merge and form a larger cluster \cite{sorrentino2016complete, Siddique2018symmetry, schaub2016graph}. 
We find the existence regime of different invariant states as described in Sec.~\ref{sec:existance_of_patt}. In the following, we derive the $\sigma$ range for the existence of patterns observed in Fig.~\ref{fig:time_series}.

\subsubsection{Van der Pol oscillators}
The activity pattern $\mathcal{P}_{10}$ of the network of Van der Pol oscillators is depicted in
Fig.~\ref{fig:time_series}(a). It is evident that active clusters in  $\mathcal{P}_{10}$ manifest as amplitude death states as $\sigma$ increases (see Figs.~\ref{fig:time_series}(b) and \ref{fig:time_series}(c)). 
We calculate the critical value of the coupling strength at which the clusters become inactive and reach $\x^*=[0 ~0]^T$ from the invariant pattern $\mathcal{P}_{10}$, which ultimately results in the network converging to complete amplitude death. In pattern $\mathcal{P}_{10}$, the evolution equations corresponding to the clusters $\{1\}$ and $\{6\}$ are given by (from Eqs.~\eqref{eq:indep_clus_dyn_2} and~\eqref{eq:vanderpol_osc})
\begin{equation}\label{eq:vanderpol_clus_1_6}
\begin{split}
\dot x &=y-  6\sigma \sigma_x x,\\
\dot y &= -x +\mu(1-x^2)y +4\sigma \sigma_y y.
\end{split}
\end{equation}
If $\{1\}$ and $\{6\}$ become inactive, we get $\dot x =\dot y =0$ and Eq.~\eqref{eq:vanderpol_clus_1_6} admits the solution $\x^*=[0 ~0]^T$.
\paragraph{stability analysis:}
The stability of the solution $[0,0]^T$ is determined by the Jacobian matrix  
\begin{equation*}\label{}
J=\begin{bmatrix} 
-6\sigma \sigma_x & 1\\
-1 & 4\sigma \sigma_y+\mu
\end{bmatrix},
\end{equation*}
from which the condition for the largest real part of the eigenvalues to be negative can be obtained as 
\begin{equation*}\label{eq:critical_coupling_vander_pol}
\sigma > \frac{\mu}{6\sigma_x-4\sigma_y},
\end{equation*}
where we have assumed that $(6\sigma_x-4\sigma_y)>0$. Similarly, for clusters $\{2,3\}$ and $\{4,5\}$, the manifestation of amplitude death occurs at $\sigma > \mu /(6\sigma_x-2\sigma_y)$, where the complete network becomes inactive due to the onset of complete amplitude death.
If $\sigma^v_1$ represents the critical coupling for the clusters $\{1\}$ and $\{6\}$ to be inactive, and $\sigma^v_2$ for the clusters $\{2,3\}$ and $\{4,5\}$, then invariant patterns of the  Van der Pol oscillators exist in the range of the coupling strength
\begin{align}\label{eq:vp_invariant_sigma_range}
\begin{split}
\sigma (\mathcal{P}_{10}) &\le \sigma^v_1,\\
\sigma (\mathcal{P}_6) &\le \sigma^v_2,\\
-\infty< \sigma (\mathcal{P}_{0}) &< \infty.
\end{split}
\end{align}
While writing Eq.~\eqref{eq:vp_invariant_sigma_range}, we have utilized the fact that the quotient dynamics in pattern $\mathcal{P}_6$ can still coexist with that in pattern $\mathcal{P}_{10}$.

\subsubsection{Stuart-Landau oscillators}
The network of Stuart-Landau oscillators also exhibits the pattern $\mathcal{P}_{10}$, as observed in Fig.~\ref{fig:time_series}(f), and the active clusters manifest as inactive clusters via the onset of the amplitude or oscillation death states as $\sigma$ varies (see Figs.~\ref{fig:time_series}(d)-\ref{fig:time_series}(g)). If clusters $\{1\}$  and $\{6\}$ become inactive in $\mathcal{P}_{10}$, then the corresponding evolution equations can be written, from Eqs.~\eqref{eq:indep_clus_dyn_2} and~\eqref{eq:sl_osc}, as 
\begin{equation}\label{eq:sl_steady_sol}
\begin{split}
    \dot x&=0=[\lambda -(x^2+y^2)] x -\omega y - \sigma \sigma_x x,\\
    \dot y&=0=[\lambda -(x^2+y^2)] y +\omega x- \sigma \sigma_y y.
\end{split}
\end{equation}
Similarly, the evolution equations for the  clusters $\{2,3\}$ and $\{4,5\}$ are the same except that $\sigma \rightarrow 2\sigma$. 
It is straightforward to  verify that  the solution $\x^*=[0~ 0]^T$  does not depend on $\sigma$, whereas the other fixed-point solution 
\begin{equation*}\label{eq:sol_osc_death_stuart_landau}
x^*= \pm \sqrt{\frac{(\lambda-\omega c_{\mp} -\sigma \sigma_x)}{1+c_{\mp}^2}},~~
y^*=  x^* c_{\mp},
\end{equation*}
where
\begin{equation*}\label{}
c_{\mp} =\frac{\sigma (\sigma_y-\sigma_x) \mp \sqrt{(\sigma(\sigma_y-\sigma_x))^2 -4 \omega^2}}{2 \omega}
\end{equation*}
exists for  $|\sigma| \ge |2\omega/(\sigma_y-\sigma_x)|$. Note that $\sigma \rightarrow \infty$ as $\sigma_x \rightarrow \sigma_y$ and is undefined if $\omega=0$. Therefore, the oscillation death manifests only when
$\sigma_x \ne \sigma_y$ and $\omega \ne 0$.

\begin{figure*}
 \includegraphics[width=\linewidth]{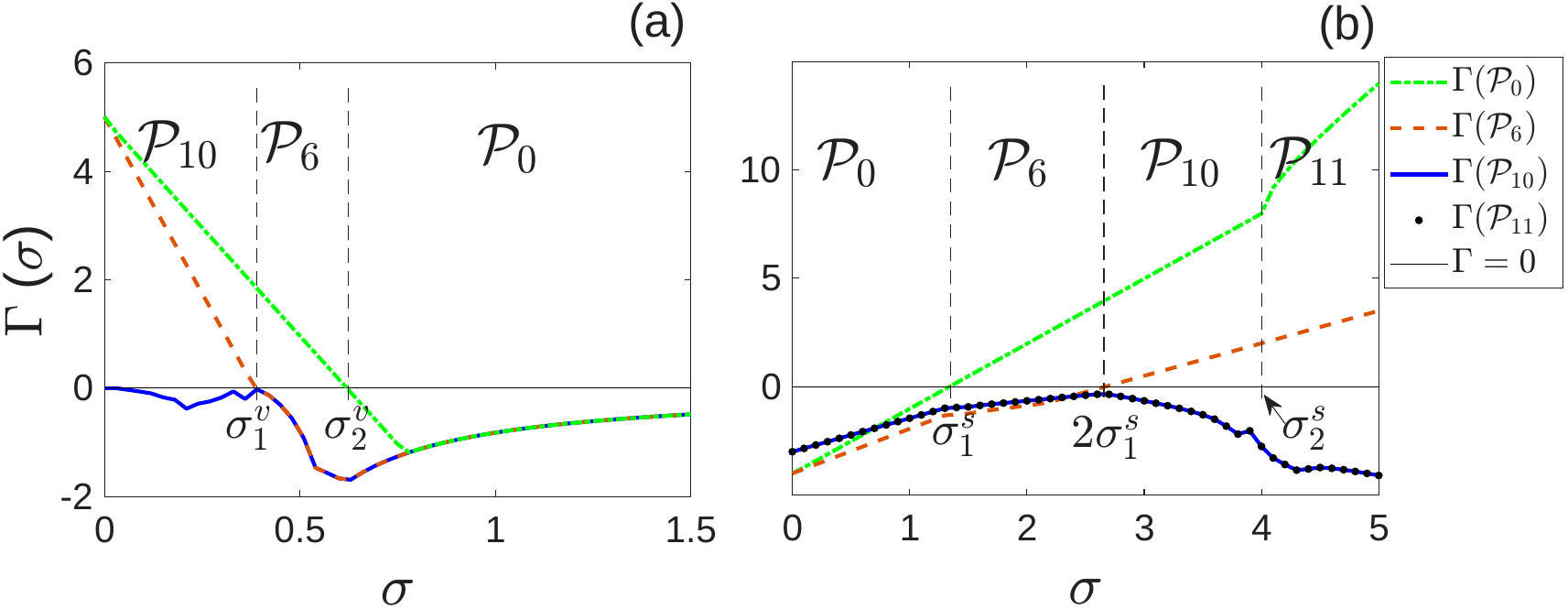}
 \caption{Stability analysis of activity patterns shown in Fig.~\ref{fig:time_series}: (a) Van der Pol oscillators and 
 (b) Stuart-Landau oscillators. A negative $\Gamma(\sigma)$ value indicates the stability of the associated invariant pattern. Each line plot corresponds to a pattern. The vertical dashed lines show stable $\sigma$ regimes of different patterns.}
 \label{fig:lyapunov_exponents}
 \end{figure*}

\paragraph{stability analysis:}

From Eq.~\eqref{eq:sl_steady_sol}, the stability of fixed point solutions of clusters $\{1\}$ and $\{6\}$ can be determined from the Jacobian matrix 
\begin{equation*}\label{}
J=\begin{bmatrix} 
\lambda-3{x^*}^2 -y{^*}^2 -\sigma \sigma_x & -2x^*y^*-\omega\\
-2x^*y^* + \omega & \lambda-{x^*}^2 -3{y^*}^2 -\sigma \sigma_y
\end{bmatrix}.
\end{equation*}
Now, eigenvalues corresponding to the fixed point $\x^*=[0~ 0]^T$ can be obtained as
\begin{equation*}\label{}
\Lambda_{1,2} =\frac{2 \lambda -\sigma(\sigma_x+\sigma_y) \pm \sqrt{\{\sigma(\sigma_x-\sigma_y)\}^2 -4 \omega^2\}}}{2}.
\end{equation*}
If $|\sigma|< |2 \omega/(\sigma_x-\sigma_y)|$, the $\max\!\big(\Re(\Lambda_1), \Re(\Lambda_2)\big)$ is negative if $(2 \lambda -\sigma(\sigma_x+\sigma_y))<0$, which for chosen parameters, determines the stable $\sigma$ regime as
\begin{equation*}\label{}
{\sigma} < \left|\frac{2 \lambda}{\sigma_x+\sigma_y}\right|.
\end{equation*}
Note that the stability conditions for the clusters $\{2,3\}$ and $\{4,5\}$ also turn out to be the same except $\sigma \rightarrow 2\sigma$. Similarly, the oscillation death solutions (those corresponding to $c_{-}$) for the clusters $\{1\}$ and $\{6\}$ are stable for
\begin{equation*}\label{}
\sigma \ge \frac{2 \omega}{\sigma_y -\sigma_x}. 
\end{equation*}
We find that other solutions corresponding to $c_{+}$ are unstable.  In summary, the invariant patterns of the  network of Stuart-Landau oscillators  exist  in the range of the coupling strength
\begin{align}\label{eq:sl_invariant_st_sigma_range}
\begin{split}
-\infty < &\sigma (\mathcal{P}_{0}) < \infty,\\
\sigma^s_1 \le &\sigma (\mathcal{P}_6) < \sigma^s_2,\\
2 \sigma^s_1 \le &\sigma (\mathcal{P}_{10}) < \sigma^s_2,\\
\sigma^s_2 \le &\sigma (\mathcal{P}_{11}) < 2\sigma^s_2,\\
2\sigma^s_2 \le &\sigma (\mathcal{P}_{12}),
\end{split}
\end{align}
where $\sigma^s_1= |\lambda/(\sigma_x+\sigma_y)|$ and $\sigma^s_2= \omega/(\sigma_y-\sigma_x)$. The invariant pattern $\mathcal{P}_{12}$ corresponds to the complete oscillation death state due to the onset of oscillation death in the clusters $\{1\}$ and $\{6\}$. 
Note that, for different clusters, heterogeneity in their degrees causes the manifestation of the inactive state at different coupling strengths.

\subsection{Stability analysis}
The largest transverse Lyapunov exponent $\Gamma(\sigma)$ corresponding to the distinct activity patterns is
depicted in Fig.~\ref{fig:lyapunov_exponents} as a function of the coupling strength. Different activity patterns
result in a distinct matrix $T'$, which is being used in \eqref{eq:perturb_dyn_xi_coordinates} to obtain the corresponding 
evolution equations for transverse perturbations and transverse Lyapunov exponents. It is obvious that $\Gamma(\sigma)<0$
corroborates the stability of an invariant pattern that exists in a particular range of $\sigma$.  
The vertical lines in Fig.~\ref{fig:lyapunov_exponents} demarcate the coupling range into regions in which distinct invariant patterns, or attractors, observed in Fig.~\ref{fig:time_series} exist and are stable. Although, we have plotted $\Gamma(\sigma)$ for the entire explored range of $\sigma$ for a smooth view of $\Gamma(\sigma)$, the largest transverse Lyapunov exponent $\Gamma(\sigma)$ in Fig.~\ref{fig:lyapunov_exponents} for a pattern should be read only when its corresponding quotient dynamics exist. 
For instance, $\Gamma$ values corresponding to $\mathcal{P}_6$ predict its stability for $\sigma\in(0, \sigma_2^v]$ only. For $\sigma >\sigma_2^v$, the quotient dynamics changes to $\mathcal{P}_{0}$ and negative $\Gamma$ values only shows that all $\deltaxi_i$ (shown in Sec.~\ref{subsec:stability inactive clusters}) decays to zero, which is valid in pattern $\mathcal{P}_{0}$ as well but should be ignored because of change of quotient dynamics.
From Eq.~\eqref{eq:vp_invariant_sigma_range} and $\Gamma(\sigma)$ values from Fig.~\ref{fig:lyapunov_exponents}(a), we conclude that invariant patterns of Van der Pol oscillators are stable for
\begin{align*}
0 < &\sigma (\mathcal{P}_{10}) \le \sigma^v_1,\\
\sigma^v_1 < &\sigma (\mathcal{P}_6) \le \sigma^v_2,\\
\sigma^v_2 < &\sigma (\mathcal{P}_{0}).
\end{align*}
Likewise, from Eq.~\eqref{eq:sl_invariant_st_sigma_range} and $\Gamma(\sigma)$ values from Fig.~\ref{fig:lyapunov_exponents}(b), invariant patterns of Stuart Landau oscillators are stable for 
\begin{align*}
0 < &\sigma (\mathcal{P}_{0}) < \sigma^s_1,\\
\sigma^s_1 \le &\sigma (\mathcal{P}_6) < 2\sigma^s_1,\\
2\sigma^s_1 \le &\sigma (\mathcal{P}_{10}) < \sigma^s_2,\\
\sigma^s_2 \le &\sigma (\mathcal{P}_{11})  < 2\sigma^s_2.\\
\end{align*}
\section{Conclusion}\label{sec:conclusions}

Investigations on pattern formation in networks of dynamical systems have largely remained focused on synchrony-based patterns ~\cite{pecora2014cluster, sorrentino2016complete, cho2017stablechimera, kumar2024symmetry, bollt2023fractal, Siddique2018symmetry}, which refers to states in which dynamical elements in a network are partitioned into groups or clusters purely based on the characteristics of synchronization, such as chimera or cluster synchronization \cite{parastesh2021chimeras}. On the contrary, inclusion of activeness of nodes in these synchrony patterns generates a wealth of patterns different from synchrony patterns and have relatively remained less explored, particularly the partial death states.  

In this work, we have elucidated the emergence of activity-based patterns consisting of active-inactive clusters of synchronized nodes in complex networks with underlying permutation symmetries. In particular, the mean-field from the active clusters should be nullified for the coexistence of active-inactive clusters in a network. 
We have shown that such coexistence can be achieved in complex networks with underlying permutation symmetry, when the internal dynamics and coupling functions are odd functions of the phase space.
The symmetries of a network generate antisynchronized clusters, facilitating zero mean-field input for some clusters to remain in an inactive state. A method to systematically generate these invariant patterns from network and quotient automorphisms is presented, where full network symmetries identify synchronized clusters and quotient symmetries further identify active-inactive state of these synchronized clusters. We presented a method to establish the existence of these activity patterns for all times, under the assumptions that the coupling functions can be decomposed into a sum of two functions and that each active cluster receives a net zero input.

We have corroborated the distinct invariant patterns, identified via the underlying symmetry of the network, numerically by employing the networks of  Van der Pol and Stuart-Landau oscillators, both exhibiting distinct clusters of identically synchronized nodes with some clusters in antisynchrony and others in amplitude or oscillation death states, leading to a rich variety of activity patterns.
Furthermore, we have shown that active clusters reach inactive states at different coupling strengths, which is determined by their degree, resulting in the network switching from one pattern to another, and ultimately converging into a complete amplitude or oscillation death state. A unified stability analysis, which identifies the stability of identical synchrony, antisynchrony, and inactive states of clusters, has been presented. Using our method, the coupling range in which numerically observed patterns are stable has been presented. 

In many complex systems, particularly in the human brain, it is desired that not all brain regions should get activated while performing different tasks or in response to external stimuli \cite{Gerdes2010brainactivations, Dedreu2019brainactivity}. Therefore, activity patterns in which different regions of a network are simultaneously in active and inactive states become relevant and may contribute to developing a deeper understanding of selective activation and functional segregation in brain-like complex networks. Our work highlights the importance of permutation symmetries in network structures and features essential in dynamical systems to achieve inactivity in parts of the network regions, while other regions remain active.

\begin{acknowledgments}
A.K. acknowledges the financial support from IISER Thiruvananthapuram. The work of VKC is supported by the computational facilities provided by the DST, New Delhi, by DST-FIST (SR/FST/PS-1/2020/135). DVS  is supported by the ANRF Project under Grant No. ANRF/ARG/2025/008542/PS.
\end{acknowledgments}

\section{Author Declarations}
\noindent
\textbf{Conflict of Interest.}
The authors declare no conflicts of interest.

\noindent
\textbf{Author Contributions.}
\textbf{Anil Kumar}: Conceptualization, Investigation, Methodology, Software, Validation, Visualization, Writing – review \& editing. \textbf{V.K. Chandrasekar}: Funding acquisition (equal), Writing – review\& editing. \textbf{D.V. Senthilkumar}: Funding acquisition (equal), Writing – review \& editing, Supervision.

\section*{Data Availability}
The data that support the findings of this study are available from the corresponding author upon reasonable request.

\bibliography{references}

@article{parastesh2021chimeras,
title = {Chimeras},
journal = {Phys. Rep.},
volume = {898},
pages = {1-114},
year = {2021},
issn = {0370-1573},
doi = {https://doi.org/10.1016/j.physrep.2020.10.003},
url = {https://www.sciencedirect.com/science/article/pii/S0370157320304014},
author = {Fatemeh Parastesh and Sajad Jafari and Hamed Azarnoush and Zahra Shahriari and Zhen Wang and Stefano Boccaletti and Matjaž Perc}
}

@article{arenas2008,
title = {Synchronization in complex networks},
journal = {Phys. Rep.},
volume = {469},
number = {3},
pages = {93-153},
year = {2008},
issn = {0370-1573},
doi = {https://doi.org/10.1016/j.physrep.2008.09.002},
url = {https://www.sciencedirect.com/science/article/pii/S0370157308003384},
author = {Alex Arenas and Albert Díaz-Guilera and Jurgen Kurths and Yamir Moreno and Changsong Zhou}
}

@article{bollt2023fractal,
  title={Fractal basins as a mechanism for the nimble brain},
  author={Bollt, Erik and Fish, Jeremie and Kumar, Anil and Roque dos Santos, Edmilson and Laurienti, Paul J},
  journal={Sci. Rep.},
  volume={13},
  number={1},
  pages={20860},
  year={2023},
  publisher={Nature Publishing Group UK London},
url={https://doi.org/10.1038/s41598-023-45664-5}
}

@article{pecora2014cluster,
  title={Cluster synchronization and isolated desynchronization in complex networks with symmetries},
  author={Pecora, Louis M and Sorrentino, Francesco and Hagerstrom, Aaron M and Murphy, Thomas E and Roy, Rajarshi},
  journal={Nat. Commun.},
  volume={5},
  number={1},
  pages={4079},
  year={2014},
  url={https://doi.org/10.1038/ncomms5079},
  publisher={Nature Publishing Group UK London}
}

@article{schaub2016graph,
    author = {Schaub, Michael T. and O'Clery, Neave and Billeh, Yazan N. and Delvenne, Jean-Charles and Lambiotte, Renaud and Barahona, Mauricio},
    title = "{Graph partitions and cluster synchronization in networks of oscillators}",
    journal = {Chaos},
    volume = {26},
    number = {9},
    pages = {094821},
    year = {2016},
    month = {08},
    issn = {1054-1500},
    doi = {10.1063/1.4961065},
    url = {https://doi.org/10.1063/1.4961065}
}

@article{poel2015partial,
  title = {Partial synchronization and partial amplitude death in mesoscale network motifs},
  author = {Poel, Winnie and Zakharova, Anna and Sch\"oll, Eckehard},
  journal = {Phys. Rev. E},
  volume = {91},
  issue = {2},
  pages = {022915},
  numpages = {12},
  year = {2015},
  month = {Feb},
  publisher = {American Physical Society},
  doi = {10.1103/PhysRevE.91.022915},
  url = {https://link.aps.org/doi/10.1103/PhysRevE.91.022915}
}

@article{saxena2012amplitude,
title = {Amplitude death: The emergence of stationarity in coupled nonlinear systems},
journal = {Phys. Rep.},
volume = {521},
number = {5},
pages = {205-228},
year = {2012},
issn = {0370-1573},
doi = {https://doi.org/10.1016/j.physrep.2012.09.003},
url = {https://www.sciencedirect.com/science/article/pii/S0370157312002645},
author = {Garima Saxena and Awadhesh Prasad and Ram Ramaswamy}
}

@article{KOSESKA2013oscillation,
title = {Oscillation quenching mechanisms: Amplitude vs. oscillation death},
journal = {Phys. Rep.},
volume = {531},
number = {4},
pages = {173-199},
year = {2013},
issn = {0370-1573},
doi = {https://doi.org/10.1016/j.physrep.2013.06.001},
url = {https://www.sciencedirect.com/science/article/pii/S0370157313002652},
author = {Aneta Koseska and Evgeny Volkov and Jürgen Kurths}
}

@article{della2020symmetries,
  title={Symmetries and cluster synchronization in multilayer networks},
  author={Della Rossa, Fabio and Pecora, Louis and Blaha, Karen and Shirin, Afroza and Klickstein, Isaac and Sorrentino, Francesco},
  journal={Nat. Commun.},
  volume={11},
  number={1},
  pages={3179},
  year={2020},
url={https://doi.org/10.1038/s41467-020-16343-0},
  publisher={Nature Publishing Group UK London}
}

@article{ZOU2021quenching,
title = {Quenching, aging, and reviving in coupled dynamical networks},
journal = {Phys. Rep.},
volume = {931},
pages = {1-72},
year = {2021},
issn = {0370-1573},
doi = {https://doi.org/10.1016/j.physrep.2021.07.004},
url = {https://www.sciencedirect.com/science/article/pii/S0370157321002799},
author = {Wei Zou and D. V. Senthilkumar and Meng Zhan and Jürgen Kurths}
}

@article{sorrentino2016complete,
author = {Francesco Sorrentino  and Louis M. Pecora  and Aaron M. Hagerstrom  and Thomas E. Murphy  and Rajarshi Roy },
title = {Complete characterization of the stability of cluster synchronization in complex dynamical networks},
journal = {Sci. Adv.},
volume = {2},
number = {4},
pages = {e1501737},
year = {2016},
doi = {10.1126/sciadv.1501737},
URL = {https://www.science.org/doi/abs/10.1126/sciadv.1501737}
}

@article{Siddique2018symmetry,
  title = {Symmetry- and input-cluster synchronization in networks},
  author = {Siddique, Abu Bakar and Pecora, Louis and Hart, Joseph D. and Sorrentino, Francesco},
  journal = {Phys. Rev. E},
  volume = {97},
  issue = {4},
  pages = {042217},
  numpages = {9},
  year = {2018},
  month = {Apr},
  publisher = {American Physical Society},
  doi = {10.1103/PhysRevE.97.042217},
  url = {https://link.aps.org/doi/10.1103/PhysRevE.97.042217}
}

@article{zakharova2014chimera,
  title = {Chimera Death: Symmetry Breaking in Dynamical Networks},
  author = {Zakharova, Anna and Kapeller, Marie and Sch\"oll, Eckehard},
  journal = {Phys. Rev. Lett.},
  volume = {112},
  issue = {15},
  pages = {154101},
  numpages = {5},
  year = {2014},
  month = {Apr},
  publisher = {American Physical Society},
  doi = {10.1103/PhysRevLett.112.154101},
  url = {https://link.aps.org/doi/10.1103/PhysRevLett.112.154101}
}

@article{rodrigues2016kuramoto,
title = {The Kuramoto model in complex networks},
journal = {Phys. Rep.},
volume = {610},
pages = {1-98},
year = {2016},
issn = {0370-1573},
doi = {https://doi.org/10.1016/j.physrep.2015.10.008},
url = {https://www.sciencedirect.com/science/article/pii/S0370157315004408},
author = {Francisco A. Rodrigues and Thomas K. DM. Peron and Peng Ji and Jürgen Kurths}
}

@article{lodi2021one,
  title={One-way dependent clusters and stability of cluster synchronization in directed networks},
  author={Lodi, Matteo and Sorrentino, Francesco and Storace, Marco},
  journal={Nat. Commun.},
  volume={12},
  number={1},
  pages={4073},
  year={2021},
  publisher={Nature Publishing Group UK London},
url={https://doi.org/10.1038/s41467-021-24363-7}
}

@article{Stewart2003symmetry,
author = {Stewart, Ian and Golubitsky, Martin and Pivato, Marcus},
title = {Symmetry Groupoids and Patterns of Synchrony in Coupled Cell Networks},
journal = {SIAM J. Appl. Math.},
volume = {2},
number = {4},
pages = {609-646},
year = {2003},
doi = {10.1137/S1111111103419896},
URL = {https://doi.org/10.1137/S1111111103419896}
}

@article{Zhang2020symmetry,
author = {Zhang, Yuanzhao and Motterqu, Adilson E.},
title = {Symmetry-Independent Stability Analysis of Synchronization Patterns},
journal = {SIAM Rev.},
volume = {62},
number = {4},
pages = {817-836},
year = {2020},
doi = {10.1137/19M127358X},
URL = {https://doi.org/10.1137/19M127358X}
}

@article{Majhi2019chimera,
title = {Chimera states in neuronal networks: A review},
journal = {Phys. Life Rev.},
volume = {28},
pages = {100-121},
year = {2019},
issn = {1571-0645},
doi = {https://doi.org/10.1016/j.plrev.2018.09.003},
url = {https://www.sciencedirect.com/science/article/pii/S1571064518301088},
author = {Soumen Majhi and Bidesh K. Bera and Dibakar Ghosh and Matjaž Perc}
}

@article{Abrams2004chimeras,
  title = {Chimera States for Coupled Oscillators},
  author = {Abrams, Daniel M. and Strogatz, Steven H.},
  journal = {Phys. Rev. Lett.},
  volume = {93},
  issue = {17},
  pages = {174102},
  numpages = {4},
  year = {2004},
  month = {Oct},
  publisher = {American Physical Society},
  doi = {10.1103/PhysRevLett.93.174102},
  url = {https://link.aps.org/doi/10.1103/PhysRevLett.93.174102}
}

@article{Bansal2019cognitive,
author = {Kanika Bansal  and Javier O. Garcia  and Steven H. Tompson  and Timothy Verstynen  and Jean M. Vettel  and Sarah F. Muldoon },
title = {Cognitive chimera states in human brain networks},
journal = {Sci. Adv.},
volume = {5},
number = {4},
pages = {eaau8535},
year = {2019},
doi = {10.1126/sciadv.aau8535},
URL = {https://www.science.org/doi/abs/10.1126/sciadv.aau8535}
}

@article{Prasad2006phase,
  title = {Phase-flip bifurcation induced by time delay},
  author = {Prasad, Awadhesh and Kurths, J\"urgen and Dana, Syamal Kumar and Ramaswamy, Ramakrishna},
  journal = {Phys. Rev. E},
  volume = {74},
  issue = {3},
  pages = {035204},
  numpages = {4},
  year = {2006},
  month = {Sep},
  publisher = {American Physical Society},
  doi = {10.1103/PhysRevE.74.035204},
  url = {https://link.aps.org/doi/10.1103/PhysRevE.74.035204}
}

@article{chowdhury2020effect,
  title = {Effect of repulsive links on frustration in attractively coupled networks},
  author = {Chowdhury, Sayantan Nag and Ghosh, Dibakar and Hens, Chittaranjan},
  journal = {Phys. Rev. E},
  volume = {101},
  issue = {2},
  pages = {022310},
  numpages = {10},
  year = {2020},
  month = {Feb},
  publisher = {American Physical Society},
  doi = {10.1103/PhysRevE.101.022310},
  url = {https://link.aps.org/doi/10.1103/PhysRevE.101.022310}
}

@article{RATTENBORG2000behavioral,
title = {Behavioral, neurophysiological and evolutionary perspectives on unihemispheric sleep},
journal = {Neurosci. \& Biobehav. Rev.},
volume = {24},
number = {8},
pages = {817-842},
year = {2000},
issn = {0149-7634},
doi = {https://doi.org/10.1016/S0149-7634(00)00039-7},
url = {https://www.sciencedirect.com/science/article/pii/S0149763400000397},
author = {N.C Rattenborg and C.J Amlaner and S.L Lima}
}

@article{cho2017stablechimera,
  title = {Stable Chimeras and Independently Synchronizable Clusters},
  author = {Cho, Young Sul and Nishikawa, Takashi and Motter, Adilson E.},
  journal = {Phys. Rev. Lett.},
  volume = {119},
  issue = {8},
  pages = {084101},
  numpages = {6},
  year = {2017},
  month = {Aug},
  publisher = {American Physical Society},
  doi = {10.1103/PhysRevLett.119.084101},
  url = {https://link.aps.org/doi/10.1103/PhysRevLett.119.084101}
}

@article{kumar2024symmetry,
    author = {Kumar, Anil and dos Santos, Edmilson Roque and Laurienti, Paul J. and Bollt, Erik},
    title = {Symmetry breaker governs synchrony patterns in neuronal inspired networks},
    journal = {Chaos},
    volume = {34},
    number = {11},
    pages = {113115},
    year = {2024},
    month = {11},
    issn = {1054-1500},
    doi = {10.1063/5.0209865}
}

@article{Berner2020birth,
  title = {Birth and Stabilization of Phase Clusters by Multiplexing of Adaptive Networks},
  author = {Berner, Rico and Sawicki, Jakub and Sch\"oll, Eckehard},
  journal = {Phys. Rev. Lett.},
  volume = {124},
  issue = {8},
  pages = {088301},
  numpages = {7},
  year = {2020},
  month = {Feb},
  publisher = {American Physical Society},
  doi = {10.1103/PhysRevLett.124.088301},
  url = {https://link.aps.org/doi/10.1103/PhysRevLett.124.088301}
}

@article{Berner2019multiclusters,
author = {Berner, Rico and Sch\"{o}ll, Eckehard and Yanchuk, Serhiy},
title = {Multiclusters in Networks of Adaptively Coupled Phase Oscillators},
journal = {SIAM J. Appl. Dyn. Syst.},
volume = {18},
number = {4},
pages = {2227-2266},
year = {2019},
doi = {10.1137/18M1210150}
}

@article{Thamizharasan2022exotic,
  title = {Exotic states induced by coevolving connection weights and phases in complex networks},
  author = {Thamizharasan, S. and Chandrasekar, V. K. and Senthilvelan, M. and Berner, Rico and Sch\"oll, Eckehard and Senthilkumar, D. V.},
  journal = {Phys. Rev. E},
  volume = {105},
  issue = {3},
  pages = {034312},
  numpages = {11},
  year = {2022},
  month = {Mar},
  publisher = {American Physical Society},
  doi = {10.1103/PhysRevE.105.034312},
  url = {https://link.aps.org/doi/10.1103/PhysRevE.105.034312}
}

@article{Berner2019hierarchical,
    author = {Berner, Rico and Fialkowski, Jan and Kasatkin, Dmitry and Nekorkin, Vladimir and Yanchuk, Serhiy and Schöll, Eckehard},
    title = {Hierarchical frequency clusters in adaptive networks of phase oscillators},
    journal = {Chaos},
    volume = {29},
    number = {10},
    pages = {103134},
    year = {2019},
    month = {10},
    doi = {10.1063/1.5097835},
    url = {https://doi.org/10.1063/1.5097835}
}

@article{Thamizharasan2024habbian,
  title = {Hebbian and anti-Hebbian adaptation-induced dynamical states in adaptive networks},
  author = {Thamizharasan, S. and Chandrasekar, V. K. and Senthilvelan, M. and Senthilkumar, D. V.},
  journal = {Phys. Rev. E},
  volume = {109},
  issue = {1},
  pages = {014221},
  numpages = {14},
  year = {2024},
  month = {Jan},
  publisher = {American Physical Society},
  doi = {10.1103/PhysRevE.109.014221},
  url = {https://link.aps.org/doi/10.1103/PhysRevE.109.014221}
}

@book{Pikovsky2001synchronization, 
title={Synchronization: A Universal Concept in Nonlinear Sciences}, 
author={Pikovsky, Arkady and Rosenblum, Michael and Kurths, Jürgen}, 
year={2001}, 
place={Cambridge}, 
publisher={Cambridge University Press}
}

@article{Totz2015phaselag,
  title = {Phase-lag synchronization in networks of coupled chemical oscillators},
  author = {Totz, Jan F. and Snari, Razan and Yengi, Desmond and Tinsley, Mark R. and Engel, Harald and Showalter, Kenneth},
  journal = {Phys. Rev. E},
  volume = {92},
  issue = {2},
  pages = {022819},
  numpages = {7},
  year = {2015},
  month = {Aug},
  publisher = {American Physical Society},
  doi = {10.1103/PhysRevE.92.022819},
  url = {https://link.aps.org/doi/10.1103/PhysRevE.92.022819}
}

@article{motter2013spontaneous,
  title={Spontaneous synchrony in power-grid networks},
  author={Motter, Adilson E and Myers, Seth A and Anghel, Marian and Nishikawa, Takashi},
  journal={Nat. Phys.},
  volume={9},
  number={3},
  pages={191--197},
  year={2013},
  publisher={Nature Publishing Group UK London},
url = {https://doi.org/10.1038/nphys2535}
}

@article{TANG2014synch_review,
title = {Synchronization in complex networks and its application – A survey of recent advances and challenges},
journal = {Annu. Rev. Control},
volume = {38},
number = {2},
pages = {184-198},
year = {2014},
issn = {1367-5788},
doi = {https://doi.org/10.1016/j.arcontrol.2014.09.003},
url = {https://www.sciencedirect.com/science/article/pii/S1367578814000376},
author = {Yang Tang and Feng Qian and Huijun Gao and Jürgen Kurths}
}

@article{BOCCALETTI2006complexnet,
title = {Complex networks: Structure and dynamics},
journal = {Phys. Rep.},
volume = {424},
number = {4},
pages = {175-308},
year = {2006},
issn = {0370-1573},
doi = {https://doi.org/10.1016/j.physrep.2005.10.009},
url = {https://www.sciencedirect.com/science/article/pii/S037015730500462X},
author = {S. Boccaletti and V. Latora and Y. Moreno and M. Chavez and D.-U. Hwang}
}

@article{Gardenes2007pathsto,
  title = {Paths to Synchronization on Complex Networks},
  author = {G\'omez-Garde\~nes, Jes\'us and Moreno, Yamir and Arenas, Alex},
  journal = {Phys. Rev. Lett.},
  volume = {98},
  issue = {3},
  pages = {034101},
  numpages = {4},
  year = {2007},
  month = {Jan},
  publisher = {American Physical Society},
  doi = {10.1103/PhysRevLett.98.034101},
  url = {https://link.aps.org/doi/10.1103/PhysRevLett.98.034101}
}

@article{glaze2016chimerastates,
    author = {Glaze, Tera A. and Lewis, Scott and Bahar, Sonya},
    title = {Chimera states in a Hodgkin-Huxley model of thermally sensitive neurons},
    journal = {Chaos},
    volume = {26},
    number = {8},
    pages = {083119},
    year = {2016},
    month = {08},
    issn = {1054-1500},
    doi = {10.1063/1.4961122},
    url = {https://doi.org/10.1063/1.4961122}
}

@ARTICLE{Gerdes2010brainactivations,
  
AUTHOR={Gerdes, Antje B. and Wieser, Matthias J. and Muehlberger, Andreas  and Weyers, Peter  and Alpers, Georg W. and Plichta, Michael M. and Breuer, Felix  and Pauli, Paul },
         
TITLE={Brain Activations to Emotional Pictures are Differentially Associated with Valence and Arousal Ratings},
        
JOURNAL={Front. Hum. Neurosci.},
        
VOLUME={4},

YEAR={2010},

URL={https://www.frontiersin.org/journals/human-neuroscience/articles/10.3389/fnhum.2010.00175},

DOI={10.3389/fnhum.2010.00175},

ISSN={1662-5161},
}

@ARTICLE{Dedreu2019brainactivity,
  
AUTHOR={de Dreu, Miek J.  and Schouwenaars, Irena T.  and Rutten, Geert-Jan M.  and Ramsey, Nick F.  and Jansma, Johan M. },
         
TITLE={Brain Activity Associated With Expected Task Difficulty},
        
JOURNAL={Front. Hum. Neurosci.},
        
VOLUME={13},

YEAR={2019},

URL={https://www.frontiersin.org/journals/human-neuroscience/articles/10.3389/fnhum.2019.00286},

DOI={10.3389/fnhum.2019.00286},

ISSN={1662-5161},
}

@manual{Stein2013Sage,
  author       = {William, Stein},
  title        = {SAGE: Software for Algebra and Geometry Experimentation},
  year         = {2013},
  url          = {http://www.sagemath.org}
}

@manual{GAP2005,
  author       = {{The GAP Group}},
  title        = {{GAP: Groups, Algorithms, and Programming, Version 4-4}},
  year         = {2005},
  url          = {http://www.gap-system.org}
}

\end{document}